\documentstyle[prd,aps,eqsecnum,preprint,tighten]{revtex}
\draft
\begin{document}
\title{Casimir energy of a compact cylinder under the condition
       $\varepsilon \mu =c^{-2}$}
\author{V.V. Nesterenko\thanks{E-mail: nestr@thsun1.jinr.ru},
I.G. Pirozhenko\thanks{E-mail: pirozhen@thsun1.jinr.ru}}
\address{Bogoliubov Laboratory of Theoretical Physics \\
Joint Institute for Nuclear Research, Dubna, 141980, Russia}
\date{\today}
\maketitle
\begin{abstract}
The Casimir energy  of an infinite compact cylinder placed in a uniform
unbounded medium is investigated under the continuity condition
for the light velocity  when crossing the interface. As a characteristic
parameter in the problem the ratio
$\xi^2=(\varepsilon_1-\varepsilon_2)^2/(\varepsilon_1+\varepsilon_2)^2=
(\mu_1-\mu_2)^2/(\mu_1+\mu_2)^2 \le 1$ is used, where $\varepsilon_1$
and $\mu_1$  are,
respectively, the permittivity and permeability of the material making
up the cylinder and $\varepsilon_2$ and $\mu_2$  are those for the surrounding
medium. It is  shown that the expansion of the Casimir energy  in powers
of this parameter begins with the term proportional to $\xi^4$.
The explicit formulas permitting us to find numerically the  Casimir
energy for any fixed value of  $\xi^2$ are obtained.
Unlike a compact ball with the same properties of the materials,
the Casimir forces in the problem under consideration are attractive.
The implication of the calculated Casimir energy in the flux tube model of
confinement is briefly discussed.
\end{abstract}
\thispagestyle{empty}
\pacs{12.20.Ds, 03.70.+k, 12.40.-y}

\section{Introduction}
The calculation of the Casimir energy   for boundary conditions
given on the surface of an  infinite cylinder has turned out to be  the most
complicated problem in this field~\cite{RM,DeRaad,BD,MNN,LNB,GRomeo}.
In Ref.~\cite{BD} an attempt was undertaken to predict the
Casimir energy of a conducting cylindrical shell treating the
cylinder as an intermediate configuration
between a sphere and two parallel plates. Taking into account that
the vacuum energies
of a conducting sphere and conducting plates have the opposite signs,
the authors
hypothesized that the Casimir energy of a cylindrical perfectly conducting
shell should be zero. However, a direct calculation~\cite{RM,DeRaad}
showed that this energy is negative as in the case of  parallel
conducting plates. This calculation was repeated
only in recent papers~\cite{MNN,LNB,GRomeo} by making use  of
comprehensive methods,
more simple but more formal at the same time.

Thus in spite of its half-century history the Casimir effect
still remains a problem where  physical intuition does not work,
and in order to reveal even the sign of the Casimir energy (i.e.\ the
direction of the Casimir forces) it is necessary to carry out  a consistent
detailed calculation.

     The account for dielectric and
magnetic properties of the media in the case of nonplanar
interface proved to be  a
very complicated problem in  calculation of the Casimir
energy~\cite{BKV}. However if the
light  velocity is constant when crossing  the interface, then the
calculation of the Casimir
energy of a compact ball~\cite{BNP} or
cylinder~\cite{MNN} is the same as that for conducting spherical
or cylindrical shells, respectively.
In such calculations the expansion of the Casimir energy in terms of the
parameter  $\xi^2=(\varepsilon_1-
\varepsilon_2)^2/(\varepsilon_1+\varepsilon_2)^2=
(\mu_1-\mu_2)^2/(\mu_1+\mu_2)^2\le 1$ is usually constructed, where
$\varepsilon_1$ and $\mu_1$ are, respectively, the permittivity and
permeability of the material making up the  ball or cylinder,
and $\varepsilon_2$, $\mu_2$  are those for the surrounding medium. The same
velocity of light, $c$, in both the media implies  that the condition
$\varepsilon _1 \mu_1=\varepsilon _2 \mu_2=c^{-2}$ is satisfied.

The Casimir energy of  a compact ball with the same speed of
light inside and outside~\cite{BNP} and the Casimir energy of a pure
dielectric ball~\cite{BM,BMM,Barton}
turned out to be of the same sign: they are  positive, and
consequently the Casimir forces are repulsive.\footnote{We use the terms
``ure dielectric ball" and ``pure dielectric cylinder"
for the corresponding nonmagnetic configurations with $\mu_1=\mu_2=1$
and $\varepsilon_1\not=\varepsilon_2 $.}
Moreover, the extrapolation of the result obtained under the condition
$\varepsilon\mu=c^{-2}$  to a pure dielectric ball gives a fairly good
prediction~\cite{BNP,BM}.

For a  compact cylinder under the condition $\varepsilon \mu=c^{-2}$
it has been found~\cite{MNN} that
the linear term in the Casimir energy expansion in powers of $\xi^2 $ vanishes.
Keeping in mind the situation with a compact
ball possessing the same speed of light inside and outside
and a pure dielectric
ball, it is tempting to check whether the Casimir energy of a
compact cylinder under the condition
$\varepsilon\mu=c^{-2}$ is close to the Casimir energy of
a pure dielectric cylinder.
However,
in the case of a dielectric cylinder a
principal difficulty arises, namely, in the
integral representation for the corresponding spectral $\zeta$-function
(or, in other words, for the sum of eigenfrequencies) it is impossible
to carry out the integration over the longitudinal momentum $k_z$.
On the other hand, in  Ref.~\cite{MNN} the Casimir energy of
a compact dielectric
cylinder was evaluated
by a direct summation of  the van der Waals interaction
between  individual fragments (molecules) of the cylinder. By
making use  of the dimensional regularization, a vanishing  value
for this energy  was obtained. It is worth noting that this procedure,
having been applied to a pure dielectric ball~\cite{MNg}, gives
the same result as the quantum field
theory approach~\cite{BMM}.
In view of all this, it is undoubtedly
interesting to  elucidate whether the vacuum energy of the electromagnetic
field for a compact cylinder with the condition
$\varepsilon\mu=c^{-2}$  vanishes exactly.
Therefore, the main goal of the present paper is, namely,
to extend the analysis made
in~\cite{MNN} up to the fourth order in $\xi $. To this accuracy  the Casimir
energy in question
turns out to be nonvanishing. Our consideration is concerned
with zero temperature theory only, and the main calculation
ignores dispersion.

The layout of the paper is as follows. In Sec.~II the first nonvanishing
term proportional
to $\xi^4$ is calculated in the expansion of the Casimir energy
of a compact cylinder in powers of $\xi ^2$ under the  condition
$\varepsilon\mu=c^{-2}$.
This term proves to be
negative, and  the Casimir forces seek to
contract the cylinder reducing its radius, unlike the  repulsive
forces acting on a compact ball under the same conditions.
In Sec.~III	 the Casimir energy in the problem at  hand
is calculated numerically
for several fixed values of the parameter $\xi^2$ without assuming the
smallness of $\xi^2$,  and the corresponding
plot is  presented. In Sec.~IV the implication of
the obtained results
in the flux tube model (hadronic string) describing
the quark dynamics inside the hadrons is considered.
In the Conclusion (Sec.~V) some general
properties of the Casimir effect are briefly discussed.

\section{Expansion of the Casimir energy in powers of $\xi^2$}

We start  with the formulas which
allow  us to construct the  expansion of the Casimir energy of
a compact infinite
cylinder, possessing the same speed of light inside and
outside, in powers of the parameter $\xi^2$.
The derivation of these formulas can be found in the  papers cited below.

When using the mode-by-mode summation method~\cite{MNN} or
the zeta function technique~\cite{LNB}  the
Casimir energy per unit length of a cylinder is
represented as a sum of partial energies
\begin{equation}
E=\sum\limits_{n=-\infty}^{+\infty}E_n,
\label{e_2_1}
\end{equation}
where
\begin{equation}
E_n=\frac{c}{4\pi a^2}\int_{0}^{\infty}
dy\, y \ln \left\{1-\xi^2 [y(I_n(y)K_n(y))']^2\right\}.
\label{e_2_2}
\end{equation}
Here the  condition
\begin{equation}
\varepsilon_1\mu_1=\varepsilon_2\mu_2=c^{-2}
\label{e_2_3}
\end{equation}
is assumed to hold, with $c$ being  the  velocity of light
inside and outside the cylinder (in units of that velocity in vacuum).
The parameter $\xi^2$ in Eq.~(\ref{e_2_2}) is defined by
the dielectric and magnetic
characteristics of the material of a cylinder and a surrounding
medium
\begin{equation}
\xi^2=\frac{(\varepsilon_1-\varepsilon_2)^2}{(\varepsilon_1+\varepsilon_2)^2}=
\frac{(\mu_1-\mu_2)^2}{(\mu_1+\mu_2)^2}.
\label{e_2_4}
\end{equation}

The representation~(\ref{e_2_1}), (\ref{e_2_2})  for the Casimir energy
is formal because the integral in Eq.~(\ref{e_2_2}) diverges logarithmically
at the upper limit,
and the sum over $n$ in Eq.~(\ref{e_2_1}) is also divergent. These difficulties
are removed by the following transformation of the sum~(\ref{e_2_1}):
\begin{eqnarray}
E&=&\sum\limits_{n=-\infty}^{+\infty}\left(E_n-E_{\infty}+E_{\infty}\right)=
\sum\limits_{n=-\infty}^{+\infty}\left(E_n-E_{\infty}\right)+
\sum_{n=-\infty}^{+\infty}E_{\infty} \nonumber\\
&=&\sum_{n=-\infty}^{\infty}\bar{E}_n+
E_{\infty}\sum_{n=-\infty}^{\infty} n^0,
\label{e_2_5}
\end{eqnarray}
where
\begin{eqnarray}
\bar{E}_n&=&E_n-E_{\infty},\quad n=0,\pm1,\pm2\dots,
\label{e_2_6}\\
E_{\infty}&=&\left.E_n\right|_{n\to\infty}=-\frac{c\xi^2}{16 \pi a^2}
\int_0^{\infty}\frac{z^5 dz}{(1+z^2)^3}\,{.}
\label{e_2_7}
\end{eqnarray}
A consistent treatment of the  product of two infinities
$E_{\infty}\cdot\sum_{n=-\infty}^{\infty}n^0$  leads to a
finite result  (see~\cite{MNN} and, especially,~\cite{LNB})
\begin{equation}
E_{\infty}\cdot \sum\limits_{n=-\infty}^{+\infty}n^0=\frac{c \xi^2}{16 \pi a^2}
\ln(2\pi).
\label{e_2_8}
\end{equation}
Thus
\begin{equation}
E=\sum\limits_{n=-\infty}^{\infty}\bar{E}_n+\frac{c \xi^2}{16 \pi a^2}{,}
\label{e_2_9}
\end{equation}
where
\begin{eqnarray}
\bar{E}_n&=&\bar{E}_{-n}=\frac{c}{4\pi a^2}\int_0^{\infty}dy\,y
\left\{\ln\left[1-\xi^2\sigma^2_n(y)\right]+
\frac{\xi^2}{4}\frac{y^4}{(n^2+y^2)^3}\right\}, \quad n=1,2,\dots,
\label{e_2_10}\\
\bar{E}_0&=&\frac{c}{4\pi a^2}\int_0^{\infty}dy\,y
\left\{\ln[1-\xi^2 \sigma_0^2(y)]+
\frac{\xi^2}{4}\frac{y^4}{(1+y^2)^3}\right\},\quad
\sigma_n(y)=y(I_n(y)K_n(y))'.
\label{e_2_11}
\end{eqnarray}

The Casimir energy~(\ref{e_2_9}) is defined correctly because the integrals  in
Eqs.\ (\ref{e_2_10}) and (\ref{e_2_11}) exist and the sum in Eq.~(\ref{e_2_9})
converges~\cite{MNN}. It is this formula that  should be expanded in
powers of
$\xi^2$. We confine ourselves    with the first two terms in this expansion
\begin{equation}
E \equiv E(\xi^2)=E^{(2)}\xi^2+E^{(4)}\xi^4 +O(\xi^6)\,{.}
\label{e_2_12}
\end{equation}
In the same way we have for $\bar{E}_n$
\begin{equation}
\bar{E}_n\equiv\bar{E}_n(\xi^2)=E^{(2)}_n\xi^2+E^{(4)}_n\xi^4 +O(\xi^6),
\quad n=0,1,2,\dots,
\label{e_2_13}
\end{equation}
where
\begin{eqnarray}
E_0^{(2)}&=&-\frac{c}{4\pi a^2}\int_0^{\infty}dy\,y \left[\sigma_0^2(y)-
\frac{y^2}{4(1+y^2)^3}\right]=\frac{c}{4\pi a^2}(-0.490878),
\label{e_2_14}\\
E_n^{(2)}&=&-\frac{c}{4\pi a^2}\int_0^{\infty}dy\,y \left[\sigma_n^2(y)-
\frac{y^2}{4(n+y^2)^3}\right], \quad n=1,2,\dots,
\label{e_2_15}\\
E_0^{(4)}&=&-\frac{c}{8\pi a^2}\int_0^{\infty}dy\,y \,\sigma_0^4(y)=
\frac{c}{4\pi a^2}(-0.0860808),
\label{e_2_16}\\
E_n^{(4)}&=&-\frac{c}{8\pi a^2}\int_0^{\infty}dy\,y \,\sigma_n^4(y),
\quad n=1,2,\dots\,.
\label{e-2-17}
\end{eqnarray}
The integrals in  Eqs.~(\ref{e_2_15}) and (\ref{e-2-17}) containing Bessel
functions can be calculated numerically only for $ n<n_0 $ with a certain
fixed value of~$n_0$. For all
the rest partial energies with $n\ge n_0$ one needs an analytic expression.
We derive such a formula using the uniform asymptotic expansion (UAE)
for the  product of the modified  Bessel functions~\cite{Ab+St}. Taking
into account all the terms up to the $n^{-6}$ order we can write
\begin{eqnarray}
\lefteqn{\ln\left\{1-\xi^2\left[y\frac{d}{dy}(I_n(ny)K_n(ny))
\right]^2\right\}=}
\nonumber\\
&=&-\xi^2\frac{y^4 t^6}{4 n^2}\left[1+\frac{t^2}{4 n^2}(3-30 \,t^2+35\, t^4)
+\frac{t^4}{4n^4}(9-256 \,t^2+1290 \,t^4-2037 \, t^6+1015 \,t^8)\right]
\nonumber\\
&&-\xi^4\frac{y^8 t^{12}}{32 n^4}\left[1+\frac{t^2}{2n^2}
(3-30\, t^2+35\, t^4)\right] - \xi^6 \frac{y^{12}t^{18}}{192\,n^6}
+ O\left ( \frac{1}{n^8}
\right ){,}
\label{e_2_18}
\end{eqnarray}
where $t=1/\sqrt{1+y^2}$.

Substituting this expression into Eq.~(\ref{e_2_10}) and integrating with
the use of the formula~\cite{GR}
\begin{eqnarray}
\label{e_2_19}
\int_0^{\infty}dy\,y^\alpha t^{\beta}=\frac{1}{\,2\,} \,
\frac{\Gamma\left(\displaystyle{\frac{\alpha+1}{2}}\right)
\Gamma\left(\displaystyle{\frac{\beta -\alpha-1}{2}}\right)}{\Gamma
\left(\displaystyle{\frac{\beta}{2}}\right)}, \\  \mbox{Re}\left (
\alpha +1
\right ) >0, \quad
\mbox{Re}\left(
\frac{\alpha -\beta +3}{2}
\right )<1 \nonumber
\end{eqnarray}
one obtains~\cite{MNN}
\begin{eqnarray}
\bar{E}_n&=&\bar{E}_n^{asymp}+O\left(\frac{1}{n^6}\right),
\label{e_2_20}\\
\bar{E}_n^{asymp}&=&\frac{c\,\xi^2}{4\pi a^2}\left(\frac{10-
3\,\xi^2}{960\, n^2}-
\frac{28224-7344\, \xi^2 + 720 \,\xi^4}{15482880 \,n^4}\right).
\label{e_2_21}
\end{eqnarray}
From here we find   the coefficients $E_n^{(2)}$ and $E_n^{(4)}$ entering
Eq.~(\ref{e_2_13})
\begin{eqnarray}
E_n^{(2)asymp}=\frac{c}{4\pi a^2}\left(\frac{1}{96\, n^2}-
\frac{7}{38040\, n^4}\right),
\label{e_2_22}\\
E_n^{(4)asymp}=-\frac{c}{4\pi a^2}\left(\frac{1}{320\, n^2}-
\frac{17}{560\cdot64\, n^4}\right).
\label{e-2-23}
\end{eqnarray}

Now by a direct numerical calculation it is necessary to estimate
the value $n=n_0$ starting from which the exact formulas~(\ref{e_2_15})
and (\ref{e-2-17}) can be substituted by the approximate ones
(\ref{e_2_22}) and (\ref{e-2-23}). In Ref.~\cite{MNN} it was shown that
when calculating $E^{(2)}$ one can
begin to use the approximate formula from $n_0=6$
\begin{eqnarray}
E^{(2)}&=&E_0^{(2)}+2\sum_{n=-1}^{5}E_n^{(2)}+
2\sum_{n=6}^{\infty}E_n^{(2)asymp}+\frac{c}{16\pi a^2}\ln (2\pi)
\nonumber\\
&=&E_0^{(2)}+2\sum_{n=-1}^{5}E_n^{(2)}+\frac{c}{4\pi a^2}
\left(\frac{1}{48}\sum_{n=6}^{\infty}\frac{1}{n^2}-
\frac{7}{19020}\sum_{n=6}^{\infty}\frac{1}{n^4}\right)+
\frac{c}{16 \pi a^2}\ln(2\pi) \nonumber\\
&=&\frac{c}{4\pi a^2}(-0.490878+0.027638+0.003778-0.000007+0.459469)
 \nonumber  \\
&=&\frac{c}{4\pi a^2}(0.000000).
\label{e_2_24}
\end{eqnarray}
This result obtained in~\cite{MNN} was interpreted there as the vanishing
of the Casimir energy of a compact cylinder under the condition (\ref{e_2_3}).
However,
as it will be shown below this is not the case.

 Table~\ref{table1} shows that when calculating the coefficient $E^{(4)}$
in Eq.\ (\ref{e_2_13}), one can also take $n_0=6$. As a result
we obtain for this coefficient
\begin{eqnarray}
E^{(4)}&=&E_0^{(4)}+2\sum\limits_{n=1}^{5}E_n^{(4)}+
2\sum\limits_{n=6}^{\infty}E_n^{(4)asymp}
 \nonumber \\
&=&E_0^{(4)}+2\sum\limits_{n=1}^{5}E_n^{(4)}-
\frac{c}{4\pi a^2}\left(\frac{1}{160}
\sum\limits_{n=6}^{\infty}\frac{1}{n^2}-\frac{17}{560\cdot32}
\sum\limits_{n=6}^{\infty}\frac{1}{n^4}\right)
\nonumber\\
&=&\frac{c}{4 \pi a^2} (-0.0860808-0.008315-0.0011334+0.0000018)
  \nonumber  \\
&=& -\frac{c}{4\pi a^2}\, 0.095528.
\label{e_2_25}
\end{eqnarray}

Thus, the Casimir energy of a compact cylinder  possessing the same speed
of light inside and outside does not vanish
and is defined up to the $\xi^4$ term
by the formula
\begin{equation}
E(\xi^2)=-\frac{c\xi^4}{4\pi a^2}\, 0.0955275=-0.007602\,\frac{c\xi^4}{a^2}.
\label{e-2-26}
\end{equation}
In contrast to the Casimir energy of a compact ball~\cite{BNP}
with the same properties
\begin{equation}
E_{ball}\simeq \frac{3}{64 a}c \xi^2 =0.046875 \,\frac{c \xi^2}{a}
\label{e_2_27}
\end{equation}
the Casimir energy of a cylinder under consideration turned out to be  negative.
Consequentially, the Casimir forces strive to contract the cylinder.
The numerical coefficient in Eq.~(\ref{e-2-26}) proved really to be small, for
example, in comparison with the analogous coefficient in Eq.~(\ref{e_2_27}).
Probably it is a manifestation of the vanishing of the Casimir
energy of a pure dielectric  cylinder noted in the Introduction.

\section{Numerical calculation of the Casimir energy for arbitrary $\xi ^2$}

Equations~(\ref{e_2_9})--(\ref{e_2_11}) obtained in the preceding
section enable one to calculate the Casimir energy
$E(\xi^2)$   numerically, without making any assumptions concerning
the smallness of the parameter $\xi^2$. Comparing the results obtained
by the exact formula (\ref{e_2_10}) and by the approximate one
(\ref{e_2_21})  we again find the value $n=n_0$ starting from
which  $\bar{E}_n^{asymp}$ reproduces $\bar {E}_n$ precisely enough.
In the general case there is its own $n_0$  for each value of $\xi^2$.
Obviously, one should expect a substantial  deviation from
Eq.~(\ref{e-2-26}) only for $\xi^2\simeq 1$. Moreover the main contribution
into the Casimir energy determined by the sum (\ref{e_2_9}) is given
by the term $\bar E_0$ which is evaluated now exactly using Eq.~(\ref{e_2_11})
without expanding in powers of $\xi^2$ as it has been done in the
preceding Section.

 The results of the calculations
accomplished in this way for  $E(\xi^2)$  are presented in Fig.~1
(solid curve). Here the Casimir energy defined by
Eq.~(\ref{e-2-26})  as a function of $\xi^2$ is also plotted
(dashed curve). When $\xi^2=1$ we get the Casimir energy of a perfectly
conducting cylindrical shell~\cite{MNN}. If we used
for its calculation the approximate
formula  (\ref{e-2-26}), we should
obtain for the dimensionless energy ${\cal E}=(4\pi a^2/c)E$ the
value $-0.0955$ instead of  $-0.1704$.  Thereby, the approximate formula
(\ref{e-2-26}) at this point gives a considerable error of $\sim 70\%$.
At the same time the analogous formula (\ref{e_2_27}) for a compact ball
at $\xi^2=1$ gives the Casimir energy of a perfectly conducting
spherical   shell with a few percent error~\cite{BNP,NP}.

\section{Implication of the calculated Casimir energy in the flux
tube model of confinement}

The constancy condition for the velocity of gluonic field when  crossing
the interface between two media is used, for example, in a dielectric
vacuum model (DVM) of quark confinement~\cite{Adler,FGK,PGM}. This model
has many elements in common with the bag models~\cite{Bag}, but
among the other differences, in DVM  there is no explicit condition of
the field vanishing outside the bag. It proves to be  important for
calculation
of the Casimir energy contribution to the hadronic mass in DVM.
The point is that in the case of boundaries with nonvanishing curvature
there happens a considerable (not full, however)  mutual cancellation
of the divergences from the contributions of internal  and
external (with respect to the boundary) regions. If only  the field confined
inside the cavity is considered, as in the bag
models~\cite{Milton-bag,BEKL,EBK},
then there is  no such a cancellation, and one has to remove some
divergences by means of renormalization of the
phenomenological parameter in the model defining the QCD vacuum energy density.

From a physical point of view the vanishing of the field or its normal
derivative precisely on the boundary is an unsatisfactory condition,
because due to quantum fluctuations it is impossible to measure
the field as accurately as desired at a certain point of the
space~\cite{LP}.

In the  DVM there is also considered a cavity that appears in the QCD vacuum
when the invariant $F_{\mu\nu}F^{\mu\nu}\sim{\bf E}^2-{\bf B}^2$ exceeds a
certain
critical value (${\bf E}$ and ${\bf B}$ are the color fields).
Inside the cavity the gluonic field can be treated as an Abelian field in view
of the asymptotic freedom in QCD. In this approach it is assumed
that in the QCD vacuum (outside the cavity) the dielectric constant
tends to zero $\varepsilon_2\to0$  while the magnetic permeability tends
to infinity $\mu_2\to\infty$ in such a way that the relativistic condition
$\varepsilon_2\mu_2=1$ holds. Inside the cavity $\varepsilon_1=\mu_1=1$.
As it was shown in the present paper for a compact cylinder and in
Ref.~\cite{BNP} for a compact ball, in calculation of the Casimir energy
the condition $\varepsilon_1\mu_1=\varepsilon_2\mu_2$
proves to be essential, and it is possible to take the limit
$\varepsilon_2\to0,\;\mu_2\to\infty$  in the resulting formula putting
$\xi^2=(\varepsilon_1-\varepsilon_2)^2/(\varepsilon_1+\varepsilon_2)^2=
(\mu_1-\mu_2)^2/(\mu_1+\mu_2)^2=1$.

Hence, in the DVM as a vacuum energy of  gluonic field one should
take the  Casimir energy of a perfectly conducting infinitely thin shell having
the shape either of a sphere, or expanded ellipsoid, or  cylinder.
In the last case we deal with the flux tube model of
confinement~\cite{FGK,Br,BN} in which a heavy quark and antiquark are
considered to be coupled through a cylindrical cavity (flux tube) in the
QCD vacuum.
Obviously, in the flux tube model of confinement the Casimir energy
of a compact cylinder calculated  under the condition
$\varepsilon_1\mu_1=\varepsilon_2\mu_2=1$ should be regarded as a quantum
correction to the classical string tension. To estimate this correction,
it is necessary  to define the value of the radius $a$ of the flux tube.
Without pretending at high accuracy we shall take  $a$ of the same order
as the critical radius $R_c$ in the hadronic string
model.\footnote{In principle the radius of the gluonic tube may be deduced
by minimizing  the linear density of a total
energy in this model, the QCD vacuum energy being considered to
be  negative~\cite{BN}. However
in this case $a$ is expressed through the phenomenological
parameter, the flux of the gluonic field, that in its turn requires
a definition.}
At the distances between the quarks smaller than $R_c$ the flux tube model
has no sense. In the Nambu-Goto string model $R_c$ is determined
by the string tension $M_0^2$~\cite{Alvarez,LN}
\begin{equation}
R_c^2=\frac{\pi}{6M_0^2}\,{.}
\label{e_4_1}
\end{equation}
Hence, we obtain the following estimation for the  Casimir energy
contribution into the string tension in the flux tube model
\begin{equation}
\left|\frac{8 E}{M_0^2}\right|=8\, \frac{7.6\cdot10^{-3}}{a^2 M_0^2}=
8\,\frac{7.6\cdot10^{-3}}{R_c^2 M_0^2}\simeq 0.1\,{.}
\label{e_4_2}
\end{equation}
The multiplier $8$ makes an account for the contribution
of the eight gluonic field components into the string tension.
Thus, unlike the conclusion made in~\cite[(1988)]{FGK}, the quantum correction
to the classical string tension, determined by the gluonic field
confined in the flux tube, turned out to be  essential $(\sim 10\%)$.
This fact should be taken into account in detailed examination of this model.

\section{Conclusion}
The Casimir energy of a compact cylinder under the condition
$\varepsilon\mu=c^{-2}$ does not vanish, but it is negative
with the absolute value increasing as $\xi^4$ for small  $\xi^2$.
The Casimir forces
seek to contract the cylinder.

The calculation of the vacuum energy  for the boundary conditions of
different geometries both with the account for the properties of the
materials and without such accounting
enables one to make the following general conclusion. In a concrete
problem the direction of the Casimir forces is determined only by the
geometry of the boundaries.  Dielectric and magnetic
properties  of the media cannot change the direction of these forces.

This conclusion is confirmed by the  calculation of the Casimir effect for
parallel conducting  plates, for a sphere and  cylinder, these boundaries
being considered in the vacuum or dividing the materials with different
dielectric and magnetic properties. Even a dilute dielectric
cylinder  mentioned above does not violate this pattern. Maybe
the Casimir forces in this case vanish in fact, but there are no
indications  that they can become repulsive.

The account for the  dispersion probably does not change
this inference. The  calculation of the Casimir energy carried out
in~\cite{NPLet} for a compact ball with $\varepsilon$
and $\mu$  dependent on the frequencies of electromagnetic
oscillations $\omega$ confirms this.
In Ref.~\cite{BrN} the Casimir forces affecting a compact cylinder
when $\varepsilon(\omega)\mu(\omega)=c^{-2}$ were investigated.
To remove  the divergences the authors introduced a double cutoff over
the frequency $\omega_0$  and over the angular momentum $n_0$.
The finite answer proved to be very involved and depended on the cutoff
parameters, but the Casimir forces are attractive as in our
consideration. However there are other points of view concerning
the role of dispersion in the Casimir effect~\cite{Candelas,BE,BS,BSS}.

\acknowledgments
This work was accomplished  with financial support of
Russian Foundation for Basic Research (Grant No.~97-01-00745).

\begin{figure}  
\caption{The dimensionless Casimir energy ${\cal E}(\xi^2)
=(4 \pi a^2/c)E(\xi^2)$ as a function of the parameter $\xi^2 $.
The solid curve
is obtained without assuming the smallness of $\xi^2 $ (the exact result);
the dashed curve  presents the approximate equation (2.26).}
\end{figure}

\begin{table}
\caption{The dimensionless coefficients
${\cal E}_n^{(4)}=(4\pi a^2/c) E_n^{(4)}$ and
${\cal E}_n^{(4)asymp} $ = $(4\pi a^2/c) E_n^{(4)asymp}$ calculated
according to Eqs.\ (2.17) and (2.23), respectively.}
\begin{tabular}{ccc}
$n$&${\cal E}_n^{(4)}$&${\cal E}_n^{(4)asymp}$\\
\tableline
1&0.002747&0.003599\\
2&0.000752&0.000811\\
3&0.000341&0.000353\\
4&0.000193&0.000197\\
5&0.000124&0.000125\\
6&0.000086&0.000087\\
\end{tabular}
\label{table1}
\end{table}
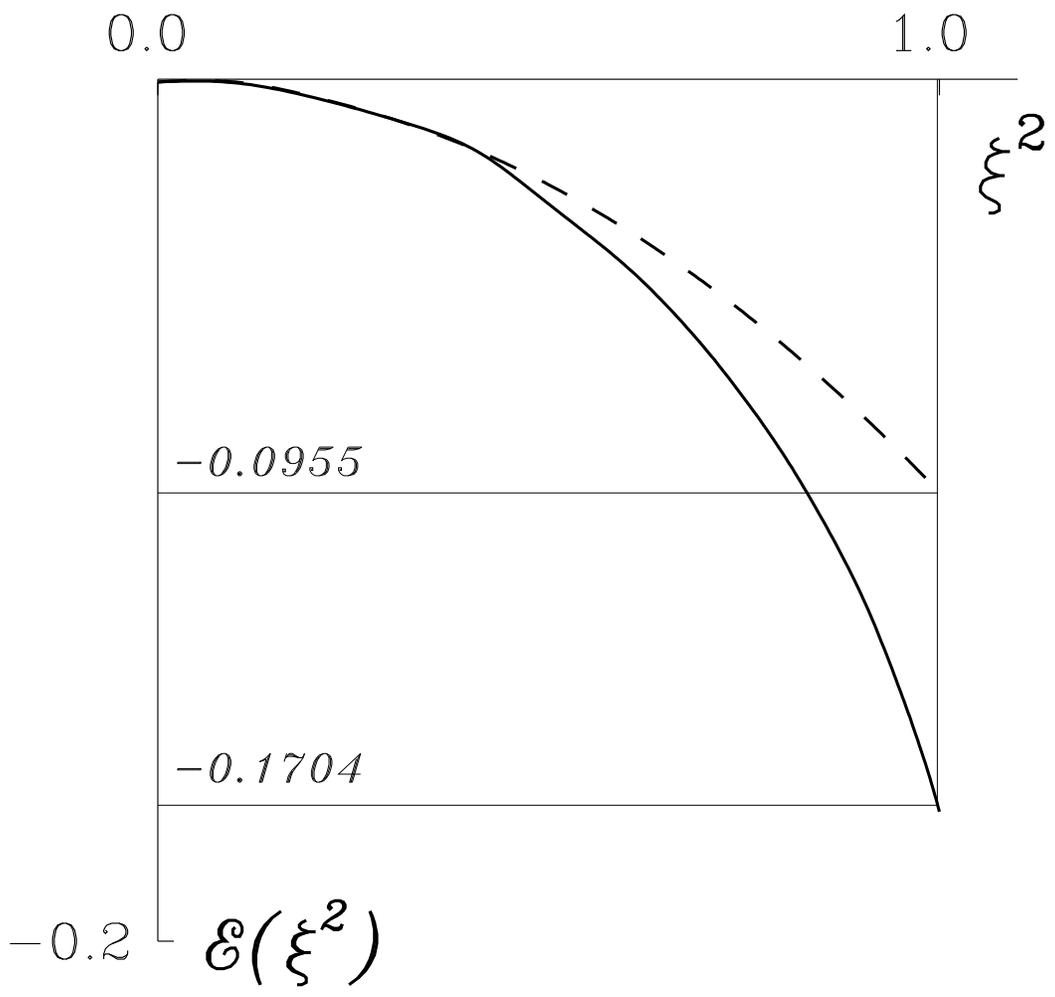
\end{document}